\documentclass[10pt,final]{IEEEtran}
\usepackage[final]{graphicx}
\usepackage{epsfig}
\usepackage{amsmath}
\usepackage{cite}

\title{On the Definition of Effective Permittivity and Permeability For Thin Composite Layers}

\author{Elena Saenz,~\IEEEmembership{Student Member,~IEEE,} Pekka M.~T.~Ikonen,~\IEEEmembership{Student Member,~IEEE,} \\
Ramon Gonzalo,~\IEEEmembership{Member,~IEEE,} and Sergei A.~Tretyakov, ~\IEEEmembership{Senior Member,~IEEE,}
\thanks{E.~Saenz and R.~Gonzalo are with the Antenna Group,
Public University of Navarra, Campus Arrosadia, E-31006 Pamplona, Spain. E-mail: elena.saenz@unavarra.es.
P.~M.~T. Ikonen and S.~A.~Tretyakov are with the Radio Laboratory/SMARAD Centre of Excellence, Helsinki
University of Technology, P.O. Box 3000, FI-02015 TKK, Finland. E-mail: pekka.ikonen@tkk.fi.} }

\begin{document}
\maketitle

\begin{abstract}
The problem of definition of effective material parameters (permittivity and permeability) for composite layers
containing only one-two parallel arrays of complex-shaped inclusions is discussed. Such structures are of high
importance for the design of novel metamaterials, where the realizable layers quite often have only one or two
layers of particles across the sample thickness. Effective parameters which describe the averaged induced
polarizations are introduced. As an explicit example, we develop an analytical model suitable for calculation of
the effective material parameters $\varepsilon_{\rm{eff}}$ and $\mu_{\rm{eff}}$ for double arrays of
electrically small electrically polarizable scatterers. Electric and magnetic dipole moments induced in the structure and the
corresponding reflection and transmission coefficients are calculated using the local field approach for the
normal plane-wave incidence, and effective parameters are introduced
through the averaged fields and polarizations. In the absence of losses both material parameters are purely real and satisfy the
Kramers-Kronig relations and the second law of thermodynamics. We compare the analytical results to the
simulated and experimental results available in the literature. The physical meaning of the
introduced parameters is discussed in
detail.
\end{abstract}


\section{Introduction}

The problem of extraction of material parameters for composite slabs implemented using complex-shape inclusions
(often referred to as metamaterial slabs) has been discussed in a large number of recent articles, see
e.g.~\cite{smi00, smi02, kos03, dep04, efr04, kos04, smi05, chen, smith_pendry, smi06} for some example
contributions. The physical meaning of many of the recent retrieval results is, however, controversial.
Typically, the $S$-parameter retrieval procedure based on the Fresnel formulae is used to extract the effective
material parameters under the normal plane-wave incidence. This technique is often known to lead to nonphysical
material parameters that violate the second law of thermodynamics: Either the real part of
$\varepsilon_{\textrm{eff}}$ or $\mu_{\textrm{eff}}$ obeys antiresonance and therefore the corresponding
imaginary part has a ``wrong'' sign, see \cite{kos03,smi05} for some results indicating this phenomenon, and
\cite{dep04,efr04,kos04} for related criticism. It is important to bear in mind that even though the
aforementioned material parameters might satisfy the Kramers-Kroning relations (thus being causal), these
parameters have no physical meaning in electromagnetic sense as physically meaningful material parameters must
satisfy simultaneously both the causality requirement and the passivity requirement  \cite{landau}. The above
discussed parameters (from now on we do not use for them the term ``material parameter'') allow, however, to
correctly reproduce the scattering results for normal plane wave incidence.

One particular example of complex slabs, referred to in the
beginning of the paper, is double arrays of small scatterers. Such
slabs present a very interesting problem as in several recent papers
the authors state that double grids of electrically short wires (or
electrically small plates) can be used to produce negative index of
refraction in the optical \cite{sha03, sha05, dol05, sha06} and in
the microwave regime \cite{zho06}. In some cases, however, the
nonphysical (antiresonant) behavior is seen in the extracted
material parameters \cite{dol05, zho06} casting doubts over the
meaningfulness of assigning negative index of refraction for these
structures.

The goal of this work is to formulate the effective material
parameters for double arrays of small scatterers through the
macroscopic polarization and magnetization, and compare the
results to those obtained using the $S$-parameter retrieval method
(the reference results are obtained by different authors
\cite{dol05}). The particles in the arrays are modeled as
electrically small electric dipoles in order to calculate, by using the
local field approach, currents induced to each of the particles
at normal plane wave incidence. Knowing the induced currents,
the reflection and transmission coefficients for the arrays, as well as
the induced electric and magnetic dipole moment densities
(macroscopic polarization and magnetization) are calculated.
Effective material parameters are then defined directly using the
expressions for the averaged fields and the macroscopic
polarization and magnetization. Finally, the double array is
represented as a slab of a homogeneous material characterized by
effective material parameters.

Since the layer is electrically thin and spatially dispersive,
it is obvious that an effective material parameter model cannot adequately and fully
describe the layer's electromagnetic properties, and, moreover,
effective parameters can be introduced in different ways.
It is important to understand the physical meaning of
different effective medium models and their applicability regions.
In this paper we compare the material
parameter extraction based on macroscopic properties calculated by
means of the local field approach with the
conventional method based on the $S$-parameter inversion.

The paper is organized in the following way: In Section II we use
the local field approach to calculate the electric dipole moments
induced in the particles and define the averaged fields needed in
the determination of the effective material parameters. The
derivation of these  parameters  is presented in Section III. In
Section IV some calculated example results are presented, and these
results are qualitatively compared with the numerical and
experimental results available in the literature. Discussion is
conducted on the physics behind the results. The work is concluded
in Section V.


\begin{figure} [b!]
    \center{\epsfig{file=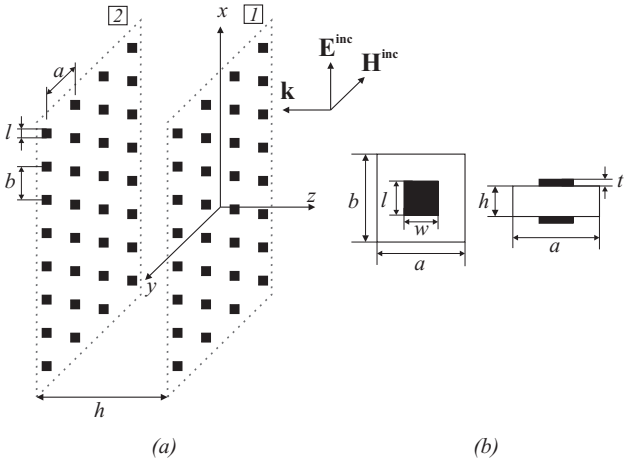,width=8.5cm}
    \caption[ ]
    {\label{geometry}
    (a) Geometry of a double array of small scatterers (square patches). (b) The unit cell.}}
\end{figure}

\section{Local field approach to determine the dipole moments and the averaged fields}

\subsection{Electric dipole moments induced in the particles}

Consider a double array of scatterers shown in
Fig.~\ref{geometry}. At this point the shape of the particles can
be arbitrary, provided that they can be modeled
using the electric dipole approximation. With this assumption the response
of each particle can be described in terms of the induced electric
dipole moment $\textbf{p}$ which is determined by the particle polarizability $\alpha$
and the local electric field at the particle position.

The array is excited by a normally incident plane expressed as
(see Fig.~\ref{geometry}(a))
\begin{equation}
    \mathbf{E}^{\rm ext} = \mathbf{E}_0e^{jk_0z} = E_0\mathbf{x}e^{jk_0z}.
    \label{Eext}
\end{equation}
The common notation for the wavenumber
$k_0=\omega\sqrt{\varepsilon_0\mu_0}$ and for the free-space wave impedance
$\eta_0=\sqrt{\mu_0/\varepsilon_0}$ is used.
The electric dipole moment induced in a reference particle sitting
in the first and the second grids reads:
\begin{equation}
    \mathbf{p}_{1,2}=\alpha_{1,2} \mathbf{E}_{1,2}^{\rm loc},
    \label{ppp}
\end{equation}
where $\mathbf{E}_{1,2}^{\rm loc}$ is the local electric field
exciting the reference particles. Since both
layers are identical in this analysis, the notation for the
particle polarizability is simplified as
$\alpha_1=\alpha_2=\alpha$. The local electric fields can be
expressed as
$$
    \mathbf{E}_1^{\rm loc}=\mathbf{E}_1^{\rm ext}+\beta(0)\mathbf{p}_1+\beta(h)\mathbf{p}_2,
$$
\begin{equation}
    \mathbf{E}_2^{\rm loc}=\mathbf{E}_2^{\rm ext}+\beta(h)\mathbf{p}_1+\beta(0)\mathbf{p}_2,
    \label{Eloc}
\end{equation}
where $\beta(0)$ is called the self-interaction coefficient and
$\beta(h)$ is the mutual interaction coefficient. Physically,
$\beta(0)$ takes into account the contribution to the local field
of the particles located in the same grid as the reference
particle, whereas $\beta(h)$ takes into account the influence of
the other grid.

By combining Eqs.~(\ref{Eext}), (\ref{ppp}) and (\ref{Eloc}), the
following system
 of equations is obtained for the unknown
dipole moments:
$$
    \alpha^{-1}\mathbf{p}_1=\mathbf{E}_0+\beta(0)\mathbf{p}_1+\beta(h)\mathbf{p}_2,
$$
\begin{equation}
    \alpha^{-1}\mathbf{p}_2=\mathbf{E}_0e^{-jk_0h}+\beta(h)\mathbf{p}_1+\beta(0)\mathbf{p}_2.
\end{equation}
Solving this system of equations, the  electric dipole
moments induced on the particles in layers 1 and 2 read:
\begin{equation}
    \mathbf{p}_1=\frac{\mathbf{E}_0}{\Delta}\bigg{[}\alpha^{-1}-\beta(0)+\beta(h)e^{-jk_0h}\bigg{]},
    \label{p1}
\end{equation}
\begin{equation}
    \mathbf{p}_2=\frac{\mathbf{E}_0}{\Delta}\bigg{[}[\alpha^{-1}-\beta(0)]e^{-jk_0h}+\beta(h)\bigg{]},
    \label{p2}
\end{equation}
where $\Delta=[\alpha^{-1}-\beta(0)]^2-\beta(h)^2$.
Approximate analytical formulas for the interaction coefficients
have been established in \cite{yat99, mas99,yat00,yat03}:
\begin{multline}
    \beta(0) =
    -\textrm{Re}\bigg{[}\frac{j\omega\eta_0}{4S_0}\bigg{(}1-\frac{1}{jk_0R}\bigg{)}e^{-jk_0R}\bigg{]}+\\
    + j \bigg{(}\frac{k_0^3}{6\pi \varepsilon_0} -
    \frac{\eta_0\omega}{2S_0}\bigg{)}, \label{beta0}
\end{multline}
\begin{multline}
    \beta(h) =
    -\textrm{Re}\bigg{[}\frac{j\omega\eta_0}{4S_0}\bigg{\{}1-\frac{1}{jk_0\sqrt{R^2+h^2}} +\\
    +\frac{h^2}{R^2+h^2}\bigg{(}1 +
    \frac{1}{jk_0\sqrt{R^2+h^2}}\bigg{)}\bigg{\}}e^{-jk_0\sqrt{R^2+h^2}}+\\
    +\frac{1}{4\pi\varepsilon_0}\bigg{\{}\frac{1}{h^3} +
    \frac{jk_0}{h^2}- \frac{k_0^2}{h}e^{-jk_0h}\bigg{]}+\\
    - j \frac{\eta_0\omega}{2S_0}\cos(k_0h), \label{betah}
\end{multline}
where $S_0$ is the unit cell area (for the rectangular-cell array $S_0=a \times b$, Fig.~\ref{geometry}(b), and
parameter $R$ is equal to $a/1.438$ \cite{mas99}). The imaginary parts of the interaction constants $\beta(0)$
and $\beta(h)$ are exact, so that the energy conservation law is satisfied. The real parts are approximate, and
the model is rather accurate for $ka < 1.5 \dots 2$ \cite{mas99,yat00,yat03}. Note that the present definition
of polarizability $\alpha$ includes also the effects of scattering, thus remaining a complex number even in the
absence of absorption. If there is no absorption in the particles,  the imaginary part of $\alpha$ reads (e.g.,
\cite{sergei}):
\begin{equation}
    \textrm{Im}\bigg{\{}\frac{1}{\alpha}\bigg{\}}=
    \frac{\eta_0\varepsilon_0\mu_0\omega^3}{6\pi} = \frac{k_0^3}{6\pi \varepsilon_0}.
    \label{im_alpha}
\end{equation}

In the far zone, the reflected field of one grid is a plane wave
field with the amplitude \cite{sergei}
\begin{equation}
    \mathbf{E}^{\rm ref}_{1,2}=-j\frac{\eta_0\omega}{2S_0}\mathbf{p}_{1,2}.
    \label{Erefamp}
\end{equation}
The scattered plane-wave field at
the reference plane $z=0$ reads:
\begin{equation}
    \mathbf{E}^{\rm ref}_{z=0}=\mathbf{E}_1^{\rm ref}+\mathbf{E}_2^{\rm
    ref}=-j\frac{\eta_0\omega}{2S_0}(\mathbf{p}_1+\mathbf{p}_2e^{-jk_0h}).
\end{equation}
Finally, the reflection and transmission coefficients can be
written as:
\begin{equation}
    R=\frac{E^{\rm ref}}{E^{\rm
    inc}}\bigg{|}_{z=0}=-j\frac{\eta_0\omega}{2S_0E_0}(p_1+p_2e^{-jk_0h}), \label{RR}
\end{equation}
\begin{equation}
    T=\frac{E^{\rm trans}}{E^{\rm inc}}\bigg{|}_{z=0}=1 -
    j\frac{\eta_0\omega}{2S_0E_0}(p_1+p_2e^{jk_0h}). \label{TT}
\end{equation}

\subsection{Definition of the averaged fields}

\begin{figure} [b!]
    \center{\epsfig{file=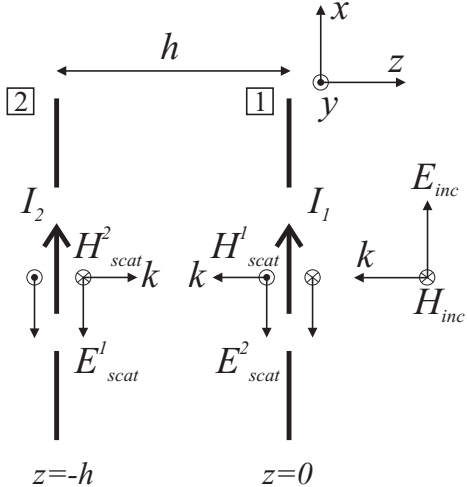,width=6.5cm}
    \caption[ ]
    {\label{IEH2}Incident and scattered fields in the problem geometry.}}
\end{figure}

The total averaged electric and magnetic fields in the vicinity of the array are, by definition, the sum of
incident field and the averaged scattered field:
\begin{equation}
\widehat{\mathbf{E}}=\langle\mathbf{E}^{\rm inc}+\mathbf{E}^{\rm ref}\rangle, \quad
\widehat{\mathbf{H}}=\langle\mathbf{H}^{\rm inc}+\mathbf{H}^{\rm ref}\rangle.
\end{equation}
Under plane-wave excitation each grid creates reflected and
transmitted plane waves having the amplitudes given by
Eq.~(\ref{Erefamp}). Those fields are the scattered fields averaged in the
transverse plane. In order to average the fields inside the
slab, the field that is scattered by the first grid and
propagates towards the second grid, and the field scattered by
the second grid and propagating towards the first grid are
considered, see Fig.~\ref{IEH2}.
The averaged electric field inside the slab can be expressed as
follows:
\begin{equation}
    \widehat{\mathbf{E}}=\frac{1}{h}\int_{-h}^0
    \bigg{[}\mathbf{E}_0e^{jk_0z}-j\frac{\eta_0\omega}{2S_0}\bigg{(}\mathbf{p}_1e^{jk_0z}+\mathbf{p}_2e^{-jk_0(z+h)}\bigg{)}\bigg{]}dz.
    \label{Eaver_int}
\end{equation}
After conducting the integration, we find the volume-averaged electric field:
\begin{equation}
    \widehat{\mathbf{E}} = e^{-jk_0h/2}\bigg{(}\mathbf{E}_0-j\frac{\eta_0\omega}{2S_0}(\mathbf{p}_1 +
    \mathbf{p}_2)\bigg{)}\frac{\sin(k_0h/2)}{k_0h/2}.
    \label{Eaver}
\end{equation}
Analogously, the averaged magnetic field can be expressed as
\begin{multline}
    \widehat{\mathbf{H}}=-\frac{e^{-jk_0h/2}}{\eta_0}\bigg{(}\mathbf{E}_0-j\frac{\eta_0\omega}{2S_0}(\mathbf{p}_1-\mathbf{p}_2)\bigg{)}\frac{\sin(k_0h/2)}{k_0h/2}.
    \label{Haver}
\end{multline}

\section{Effective material parameters}

Let us start from the common definition of the
constitutive parameters of an isotropic homogeneous material:
\begin{equation}
    \mathbf{D} = \varepsilon_0\widehat{\mathbf{E}}+
    \mathbf{P}=\varepsilon_0\varepsilon_{\textrm{eff}}\widehat{\mathbf{E}},  \quad \mathbf{P}=\frac{\mathbf{p}}{V},
    \label{D}
\end{equation}
\begin{equation}
    \mathbf{B} = \mu_0\widehat{\mathbf{H}}+
    \mathbf{M}=\mu_0\mu_{\textrm{eff}}\widehat{\mathbf{H}}, \quad \mathbf{M}=\frac{\mathbf{m}}{V}.
    \label{D}
\end{equation}
Here $\mathbf{P}$ and $\mathbf{M}$ are the volume-averaged polarization and magnetization, $\mathbf{p}$ and
$\mathbf{m}$ are the electric and magnetic dipole moments induced in the unit cell, and $V$ is the volume of the
unit cell. Knowing $\widehat{\mathbf{E}}$, $\widehat{\mathbf{H}}$, $\mathbf{P}$, and $\mathbf{M}$ for
our particular layer we {\itshape define} effective
material parameters as:
\begin{equation}
    \varepsilon_{\textrm{eff}} = 1 + \frac{\mathbf{P}}{\varepsilon_0\widehat{\mathbf{E}}}, \quad
    \mu_{\textrm{eff}} = 1 + \frac{\mathbf{M}}{\mu_0\widehat{\mathbf{H}}}.
    \label{eff}
\end{equation}
Naturally, these parameters have more limited physical meaning than the
usual parameters of a bulk homogeneous sample, but they can be used as a
measure for averaged polarizations in a thin layer. Let us next find these
parameters for our particular example grid in terms of its geometry and
inclusion polarizabilities.

In order to simplify the forthcoming notations, the sum and
difference of the electric dipole moments with the use of Eqs.~(\ref{p1}), (\ref{p2}) can be written in the
following manner:
\begin{equation}
    \mathbf{p_1}+\mathbf{p_2} = \mathbf{E}_0\frac{2e^{-jk_0h/2}\cos(k_0h/2)}{\alpha'^{-1}},
    \label{p1+p2}
\end{equation}

\begin{equation}
    \mathbf{p_1}-\mathbf{p_2} = \mathbf{E}_0\frac{2je^{-jk_0h/2}\sin(k_0h/2)}{\alpha''^{-1},}
    \label{p1-p2}
\end{equation}
where
\begin{equation}
\alpha'^{-1}=\alpha^{-1}-\beta(0)-\beta(h), \label{a1}
\end{equation}
\begin{equation}
\alpha''^{-1}=\alpha^{-1}-\beta(0)+\beta(h). \label{a2}
\end{equation}
It is important to note that for lossless particles the imaginary parts of
$\alpha'^{-1}$ and $\alpha''^{-1}$ can be solved exactly from the power balance requirement \cite{yat00}:
\begin{equation}
    \textrm{Im}\{\alpha'^{-1}\}=\frac{\eta_0\omega}{S_0}\cos^2(k_0h/2), \label{alpha'}
\end{equation}
\begin{equation}
    \textrm{Im}\{\alpha''^{-1}\}=\frac{\eta_0\omega}{S_0}\sin^2(k_0h/2). \label{alpha''}
\end{equation}
The use of exact expressions for the imaginary parts is very important:
Approximate relations would lead to complex
material parameters of lossless grids, where the imaginary part would have no
physical meaning.

The total averaged electric polarization which we need to substitute in Eq.~(\ref{eff}) reads
\begin{equation}
\mathbf{P}=\frac{\mathbf{p}_1+\mathbf{p}_2}{V}. \label{Ptot}
\end{equation}
In the following, we consider lossless
particles with the inverse polarizabilities
\begin{equation}
\alpha'^{-1} = {\rm{Re}}\{\alpha'^{-1}\} + j{\rm{Im}}\{\alpha'^{-1}\},
\end{equation}
\begin{equation}
\alpha''^{-1} = {\rm{Re}}\{\alpha''^{-1}\} + j{\rm{Im}}\{\alpha''^{-1}\}.
\end{equation}
The imaginary parts in the above relations satisfy (\ref{alpha'}) and (\ref{alpha''}).
Substituting first (\ref{p1+p2}) into (\ref{Ptot}), and then
(\ref{Ptot}) into (\ref{eff}), the effective permittivity obtained
after some mathematical manipulations is as follows:
\begin{equation}
    \varepsilon_{\textrm{eff}} =  1+\frac{k_0h}{V\varepsilon_0} \bigg{[}\textrm{Re}\{\alpha'^{-1}\}\tan\bigg{(}\frac{k_0h}{2}\bigg{)} -
    \frac{\eta_0\omega}{S_0}\sin^2\bigg{(}\frac{k_0h}{2}\bigg{)}\bigg{]}^{-1}.
    \label{final_eps_eff}
\end{equation}
Notice that the permittivity is purely real, as it
should be since the particles in the arrays are lossless.

For our system of two identical grids of small particles the
effective permeability
$\mu_{\textrm{eff}}$ defined by   (\ref{eff}) can be also expressed in terms of the induced electric dipole moments.
Considering one unit cell formed by two particles of length $l$ with $x$-directed dipole moments
$p_{1,2}$, we can write for the currents on the particles (averaged along $x$)
$I_{1,2}=j\omega p_{1,2}/l$. The magnetic moment of this pair of particles
(referred to the unit cell center) is then
\begin{equation}
m=\mu_0 l\frac{h}{2}(I_1-I_2)=j\omega\mu_0
    \frac{h}{2}(p_1-p_2). \label{mm}
\end{equation}

After inserting first (\ref{p1-p2}) into (\ref{mm}), and then
(\ref{mm}) into (\ref{eff}), the effective permeability reads:
\begin{equation}
    \mu_{\textrm{eff}}= 1+\frac{\eta_0\omega k_0 h^2}{2 V} \bigg{[}\textrm{Re}\{\alpha''^{-1}\}+
    \frac{\eta_0\omega}{2S_0}\sin(k_0h)\bigg{]}^{-1}.
    \label{final_mu_eff}
\end{equation}
We can again observe that this quantity is purely real, as it
should be in the absence of losses.
Notice that the procedure presented here can easily be generalized
to different particle geometries.

As is apparent from the definition, these effective parameters measure
cell-averaged electric and magnetic polarizations in the layer.
Can they be used to calculate the reflection and transmission coefficients
from the layer? This question will be considered next using a
numerical example.

\section{Numerical examples}

In this section, the double array of square patches studied in \cite{dol05} is
considered as a representative example of comparison between
the effective parameters introduced here and material parameters
formally extracted from measured or calculated $S$-parameters.

The geometry of the double-grid array is the same as considered above (Fig.~\ref{geometry}).  The unit cell is
characterized by the following parameters: Edge lengths of the patches $t=l$, the lattice constants $a=b$, the metal
layer thickness $p$, and the distance between the layers equals $h$. The dimensions of the unit cell considered in
\cite{dol05} are the following: $w=l=300$ nm, $t=20$ nm, $a=b=650$ nm, $p=20$ nm, $h=80$ nm. Since the double
array considered in \cite{dol05} is targeted to the use at THz frequencies, the metal thickness $p$ becomes
comparable to the dielectric spacer thickness $h$. The physical thickness of the slab is $D = h+2p$. However, at
THz frequencies one must take into account nonzero fields inside metal particles (here gold is considered, which
at these high frequencies is usually characterized by the Drude model (e.g., \cite{landau})). For this reason an
effective thickness of $D_{\rm eff} = 1.5D$ is chosen for the analysis, but we should bear in mind that this is
only a rough estimation, and an accurate determination of this effective slab thickness is very difficult.

Notice that, as it is stated in \cite{yat03}, the limitation $ka <
1.5 ... 2$ in the analytical model for the interaction constants
(Section II) is not critical for these calculations until the
lattice resonance at $ka = 2\pi$ is reached, since the reflection
properties near resonances are mainly determined by the particle
resonances. In this case, it means that the model is accurate
enough up to  approximately 460 THz.

The following rough estimates are used for calculating the particle polarizabilities. A known \cite{sergei}
antenna model for short strip particles is used by setting the particle width equal to the particle length. The
polarizability is resonant and can be characterized in the microwave regime by an equivalent LC-circuit, where $C$ is
approximated as the input capacitance of a small dipole antenna having length $l$ and radius $r_0=w/4$, and $L$
is the inductance of a short strip having length $l$, width $w$, and thickness $t$ \cite{Caulton}:
\begin{equation}
    C \approx \frac{\pi(l/2)\varepsilon_0}{\log(l/r_0)},
    \label{C}
\end{equation}
\begin{equation}
    L\approx0.2l\bigg{(}\log\frac{l}{w + t} + 1.19 + 0.22\frac{w + t}{l}\bigg{)}.
    \label{L}
\end{equation}
(The values of $l,w,t$ are inserted in eq.~(\ref{L}) in millimeters. The result is in nano-Henrys.) As mentioned
above, in the THz range the behavior of metals is different from the microwave regime. Following the treatment
in \cite{tre07}, the penetration of fields inside the particles can be represented by an additional parallel
capacitance $C_{\textrm{add}}$ and a series inductance $L_{\textrm{add}}$ \cite{tre07}:
\begin{equation}
    C_{\textrm{add}} = \frac{\varepsilon_0 w t}{l_{\textrm{eff}}},
    \label{Cadd}
\end{equation}
\begin{equation}
    L_{\textrm{add}} = \frac{l_{\textrm{eff}}}{\varepsilon_0 w t\omega_p^2},
    \label{Ladd}
\end{equation}
where $\omega_p$ is the plasma frequency of considered metal (gold
in this case) and $l_{\textrm{eff}}$ is the effective particle
length. The physical clarification of the above parameters is
available in \cite{tre07}. Due to the cosine current distribution
the effective length is approximated as $l_{\textrm{eff}}=2l$.
Finally, the particle polarizability reads:
\begin{equation}
    \alpha = \frac{(C+C_{\textrm{add}})l^2}{1-\omega^2(L+L_{\textrm{add}})(C+C_{\textrm{add}})}
\end{equation}

\begin{figure} [b!]
    \center{\epsfig{file=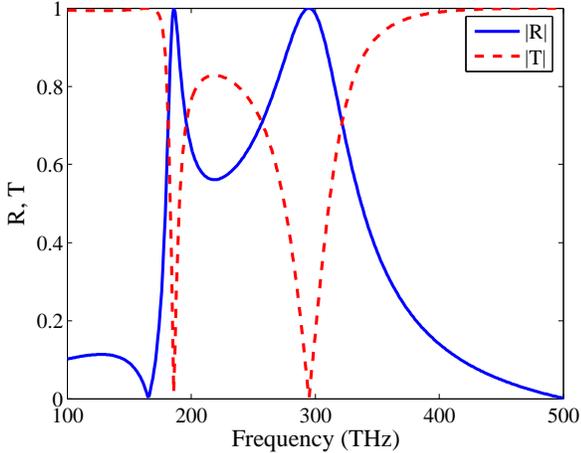,width=8.5cm}}
    \caption{Calculated reflection and transmission coefficients.{\label{RT_model_patches}}}
\end{figure}

Fig.~\ref{RT_model_patches} shows the magnitudes of the reflection
and transmission coefficients calculated using Eqs.~(\ref{RR}) and
(\ref{TT}). Comparing this result with the experimental results
presented in Fig.~4a of Ref.~\cite{dol05}, good agreement is
observed. The shape of the reflection and transmission functions
follows the ones presented in \cite{dol05}. The differences
in the amplitude levels and transmission frequencies are well
expected: The model used here does not take losses in the
particles into account.

\begin{figure} [b!]
    \center{\epsfig{file=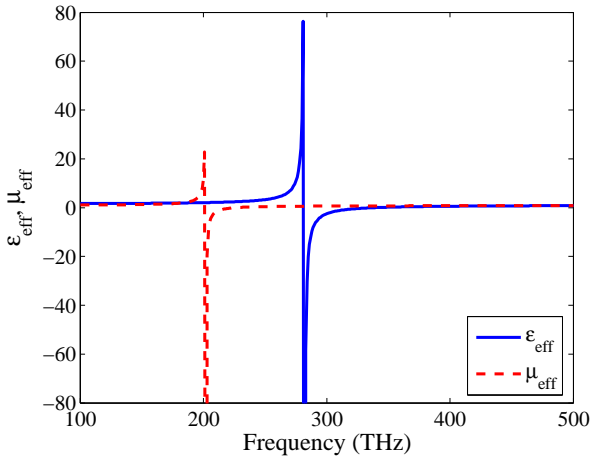,width=8.5cm}}
    \caption{Calculated effective permittivity and permeability.{\label{eps_mu_patches}}}
\end{figure}

The effective material parameters calculated using
Eqs.~(\ref{final_eps_eff}) and (\ref{final_mu_eff}) are depicted
in Fig.~\ref{eps_mu_patches}. Both material parameters clearly
behave in a physically sound manner: They are purely real, since
lossless particles have been considered, and they are growing functions
outside the resonant region. Contrary, the material parameters
extracted in \cite{dol05} for the same structure behave in a
nonphysical manner: The imaginary parts of permittivity and
permeability have the opposite signs over certain frequency ranges,
thus, the passivity requirement is violated
\cite{dep04,efr04,landau}. It is evident also from Fig.~4c and 4d
of Ref.~\cite{dol05} that the extracted material parameters do not
satisfy the Kramers-Kronig relations (e.g.,~the antiresonance in
the real part of permittivity cannot correspond to a positive imaginary
part (with the authors' time dependence assumption) if inserted into
Kramers-Kronig relations), thus, they have quite limited physical meaning
\cite{landau}. Both of these problems are avoided with the method
described here. On the other hand, the material parameters
extracted from the reflection and transmission coefficients
reproduce these properties of the layer {\itshape exactly},
while the parameters introduced here do not necessarily give so accurate
predictions for reflection and transmission coefficients.

\begin{figure} [t!]
    \center{\epsfig{file=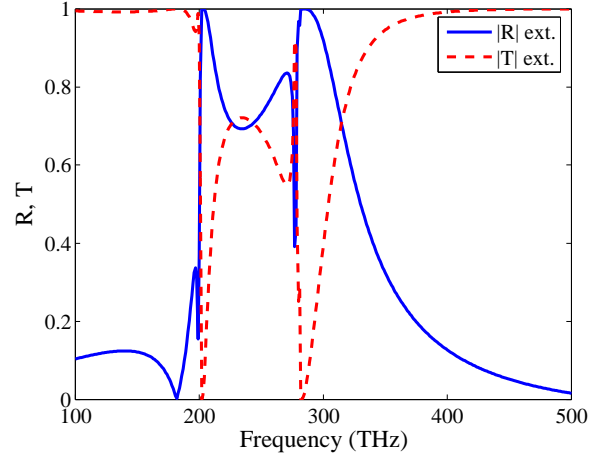,width=8.5cm}}
    \caption{Reflection and transmission coefficients calculated when the double array is represented
    as a slab of homogeneous material characterized by $\varepsilon_{\textrm{eff}},\mu_{\textrm{eff}}$
    depicted in Fig.~\ref{eps_mu_patches}.{\label{RT_ext_patches}}}
\end{figure}

This is illustrated next by  representing the double layer of particles  as a slab of a homogeneous material
having thickness $D_{\rm eff}$ and characterized by the material parameters shown in Fig.~\ref{eps_mu_patches}.
The standard transmission-line equations are used to calculate the reflection and transmission coefficients for
a normally incident plane wave. The result is depicted in Fig.~\ref{RT_ext_patches}. When comparing
Figs.~\ref{RT_model_patches} and \ref{RT_ext_patches} it can be observed that the results do not correspond
exactly to each other, even thought the principal features such as reflection frequencies around 200 and 300 THz
seen in Fig.~\ref{RT_model_patches} are reproduced in Fig.~\ref{RT_ext_patches}. This is an expected result:
There is no such uniform material (with physically sound material parameters) that would behave exactly as the
actual double grid of resonant particles. In a truly homogeneous material the unit cell over which the averaging
is performed should contain a large number of ``molecules'' yet remaining small compared to the wavelength. It
is clear that for the considered double array this is not the case, and also spatial dispersion effects cannot
be neglected.

In particular, very close to the grid resonances the induced dipole moments are very large, which corresponds to
large values of the present effective parameters and to large electrical thickness of the homogenized slab.
These large values correctly describe large polarizations in the grids, but they fail to correctly predict the
reflection and transmission coefficients. On the contrary, the effective parameters retrieved from the layer's
$S$- parameters, correctly describe these reflection and transmission coefficients, but they do not describe
polarization properties of the actual structure.

It is also important to note that the standard $S$-parameter
retrieval procedure often leads to nonphysical material
parameters (see the discussion above). The material parameters defined through
macroscopic polarization and magnetization are always physically
sound since they are defined directly by the actual dipole moments
induced in the microstructure. Thus, the effective material parameters
assigned in this way always make physical sense. 

\section{Conclusion}

In this paper an analytical model to assign effective material
parameters $\varepsilon_{\textrm{eff}}$ and $\mu_{\textrm{eff}}$
for double arrays of electrically small scatterers has been
presented. The induced electric dipole moments have been
calculated using the local field approach for a normally incident
plane wave, and the corresponding averaged fields have been
determined. The effective material parameters have been defined
directly through the macroscopic polarization and magnetization.
The derived expressions have been validated by comparing the
calculations to the numerical and experimental results available
in the literature.
The main conclusion from this study is that simple effective medium models of electrically
thin layers with complex internal structures are always limited in their
applicability and in their physical meaning. Some models are best suitable to
describe reflection and transmission coefficients at normal incidence
(like the parameters conventionally retrieved from S-parameters), other models
describe well the averaged induced polarizations in the structure and
allow one to make conclusions about, for instance, negative permeability
property (like the parameters introduced in this paper).
In applications, it is important to understand what model is used and
what properties of the layer this particular model actually describes.
Finally, it is necessary to stress that the present study has been restricted
to a particular special case of a dual layer of planar electrically
polarizable particles. This approach needs appropriate modifications and
extensions if, for instance, inclusions are also magnetically polarizable.
For slabs containing more than two layers, the direct extension of this model
corresponds to averaging over a unit cell containing three or more particles,
which apparently is not useful for electrically thick slabs.


\section*{Acknowledgements}

The research presented in this paper has been financially
supported by METAMORPHOSE NoE funded by E.C. under contract
NMP3-CT-2004-50252, CIMO Fellowship grant number TM-06-4350 and
Spanish Government under project TEC2006-13248-C04-03/TCM. The authors would like to thank
Prof. C. Simovski for his helpful comments.

\end{document}